\documentclass[aps,preprint,superscriptaddress,prb]{revtex4-2}
\usepackage{amsmath}
\usepackage{amssymb}
\usepackage{graphicx}
\usepackage[english]{babel}
\usepackage{array}
\usepackage{mathrsfs}
\usepackage{color}
\usepackage[breaklinks]{hyperref}

\newcommand{\floor}[1]{\left\lfloor{#1}\right\rfloor}

\begin{document}

\title{Revisiting the Maxwell multipoles for vectorized angular functions}

\author{Matthew Houtput}
\affiliation{Theory of Quantum and Complex Systems, Universiteit Antwerpen, B-2000 Antwerpen, Belgium}

\author{Jacques Tempere}
\affiliation{Theory of Quantum and Complex Systems, Universiteit Antwerpen, B-2000 Antwerpen, Belgium}

\begin{abstract}

Across many areas of physics, multipole expansions are used to simplify problems, solve differential equations, calculate integrals, and process experimental data. 
Spherical harmonics are the commonly used basis functions for a multipole expansion. However, they are not the preferred basis when the expression to be expanded is written as an explicit function of the unit vector on the sphere. 
Here, we revisit a different set of basis functions that are well-suited for multipole expansions of such vectorized angular functions. These basis functions are known in the literature by a variety of different names, including Maxwell multipoles, harmonic tensors, symmetric trace-free (STF) tensors, and Sachs-Pirani harmonics, but they do not seem be well-known among physicists.

We provide a novel derivation of the Maxwell multipoles that highlights their analogy with the Legendre polynomials. We also rederive their key properties, and compile a list of further properties for pratical use, since this list seems to be missing in the current literature. We show the equivalence between the Maxwell multipoles and the spherical harmonics, derive conversion formulas between the two, and motivate when it is preferable to use either formalism. Since vectorized functions occur naturally in many physical problems, we expect that the method presented in this article can simplify calculations for physicists in many different fields.
\end{abstract}

\maketitle

\section{Introduction}

The spherical harmonics $Y_{\ell,m}(\theta,\varphi)$ are an essential tool for any physical problem with spherical symmetry. Historically, the spherical harmonics have arisen from the study of differential equations \cite{ArfkenWeber}: a typical example is the Schr\"odinger equation for the hydrogen atom. They are also used extensively in the context of multipole expansion, which finds applications in virtually every physical field, from classical electromagnetism \cite{Maxwell1891, Mie1908} to modern cosmology \cite{Thorne1980, Copi2004, Weeks2004}. However, for functions in vectorized or tensorial form, the dependence on $\theta$ and $\varphi$ is obscured, and multipole expansion in the basis of spherical harmonics becomes unnecessarily complicated. In this article, we propose a solution to this problem by grouping the spherical harmonics together into tensors. This creates a new complete set of orthogonal basis functions for multipole expansion, known as Maxwell multipoles.

Similar basis functions have already been established in the literature, but do not seem to be well-known. They were first studied by Maxwell \cite{Maxwell1891} and Sylvester \cite{Sylvester1876}, and are nowadays mostly used in the context of general relativity under the name ``symmetric trace-free (STF) tensors'' \cite{Bruno2018,Thorne1980}. However, many authors use quite different conventions and even different names, including Maxwell multipoles \cite{Zou2003,Dennis2004}, tensor harmonics \cite{Efimov1979,Applequist1989}, and Sachs-Pirani harmonics \cite{Sachs1961,Pirani1965,Thorne1980}; some authors \cite{Copi2004,Rubinstein2015} do not give them a name at all. We will use the name Maxwell multipoles throughout this article. To the best of our knowledge, many of the properties we present in this article have not been discussed in the literature before.

Usually, the spherical harmonics are well suited for multipole expansion. However, there are exceptions. For example, consider the following vectorized angular function, which occurs in the expression for the electric potential of a quadrupole:
\begin{equation} \label{nFunction}
f(\theta,\varphi) = \mathbf{n}\cdot \mathcal{Q} \cdot \mathbf{n} = n_i Q_{ij} n_j,
\end{equation}
where $\mathcal{Q}$ is the quadrupole matrix (a $3 \times 3$ symmetric matrix), $\mathbf{n} := \mathbf{n}(\theta,\varphi)$ is a unit vector in the direction of $\theta$ and $\varphi$:
\begin{align}
n_x(\theta,\varphi) & := \sin\theta \cos\varphi, \label{nxDef} \\
n_y(\theta,\varphi) & := \sin\theta \sin\varphi, \label{nyDef} \\
n_z(\theta,\varphi) & := \cos\theta, \label{nzDef}
\end{align}
and we employ the Einstein convention of summation over repeated indices. If we develop this function in a multipole expansion and use the spherical harmonics as our basis, we would have to consider every component of $\mathcal{Q}$ separately, leading to six distinct terms. As we shall show in section \ref{Sec:Properties}, this problem is avoided if we use Maxwell multipoles as our basis functions instead.

The goal of this article is threefold. First, we show that the Maxwell multipoles can significantly simplify multipole expansions and integral calculations whenever the integrand explicitly depends on $\mathbf{n}$, as in \eqref{nFunction}. Therefore, the scope of Maxwell multipoles and STF tensors extends far beyond the domain of cosmology and general relativity. Second, we aim to highlight the analogy between the Maxwell multipoles and the Legendre polynomials. We present a novel derivation of the Maxwell multipoles that stresses this analogy. Third, we want this article to function as an introduction to Maxwell multipoles for those who are not familiar with them. To make the article as intuitive as possible, we build our arguments from scratch and introduce the Maxwell multipoles without referring to earlier work. Additionally, we attempt to keep our notation as simple as possible, since the tensor notation found in the literature can be quite daunting at first.

The remainder of this article is structured as follows. In section \ref{Sec:Notation}, we discuss our tensor notation and introduce the symmetric detracing operator. In section \ref{Sec:MaxwellMultipolesDef}, we introduce the Maxwell multipoles by analogy with the Legendre polynomials. In section \ref{Sec:Properties}, we list the properties of the Maxwell multipoles, and give an explicit example of how the Maxwell multipoles can simplify calculations in practice. In section \ref{Sec:SphericalHarmonics}, we derive an explicit link between the Maxwell multipoles and the more standard spherical harmonics. We summarize our results in section \ref{Sec:Conclusions}.

\section{Tensor notation} \label{Sec:Notation}
First, we summarize the notation used throughout the article. As we will see soon, the Maxwell multipoles are tensors that are both symmetric and traceless. We will write tensors using index notation: in the scope of this article, a tensor of rank $\ell$ is simply an object with $\ell$ indices, where we make no distinction between upper or lower indices. We will write a generic tensor with $\ell$ indices as $T_{i_1 i_2 \ldots i_{\ell}}$, but when the number of indices is fixed, we will write $T_{ij}$, $T_{ijk}$, and so on. Indices can take on three values: $x, y, z$.

A well-known and important concept in our context is the symmetrization operator. It extracts the fully symmetric part of a tensor, and is usually written with round brackets $()$. A similar operator is the symmetric detracing operator, which extracts the traceless symmetric part of the tensor. It is obtained by applying the symmetrisation operator to a tensor, and then removing all the traces from the result. There is no standard notation for this operator; in this article, we denote it with curly braces $\{\}$, to highlight the analogy with the symmetrisation operator. For two, three, and four indices, the result is:
\begin{align}
T_{\{ij\}} & = T_{(ij)} - \frac{1}{3} T_{aa} \delta_{ij} \label{detracing2} \\
T_{\{ijk\}} & = T_{(ijk)} - \frac{1}{5} T_{(aa i)} \delta_{jk} - \frac{1}{5} T_{(aa j)} \delta_{ik} - \frac{1}{5} T_{(aa k)} \delta_{ij} \\
T_{\{ijkl\}} & = T_{(ijkl)} - \frac{1}{7} \left( T_{(aa ij)} \delta_{kl} + 5\text{ terms that symmetrize }ijkl \right) + \frac{3}{35} T_{(aabb)} \delta_{(ij} \delta_{kl)}
\end{align}
The general formula for symmetric detracing is quite complicated and is presented in appendix \ref{App:detracing}.

Similarly to the symmerisation operator, the symmetric detracing operator satisfies the following important property:
\begin{equation}
A_{i_1 i_2 \ldots i_{\ell}} B_{\{i_1 i_2 \ldots i_{\ell}\}} = A_{\{i_1 i_2 \ldots i_{\ell}\}}B_{\{i_1 i_2 \ldots i_{\ell}\}}
\end{equation}
Because of this property, it is often possible to avoid explicitly performing the detracing until the very end of the calculation. Another interesting property is that the Kronecker delta has no traceless symmetric part:
\begin{equation} \label{TracelessDelta}
\delta_{\{ij\}} = 0
\end{equation}
which follows directly from equation \eqref{detracing2}. Therefore, the traceless symmetric part of any tensor that contains a Kronecker delta is zero.

\section{Definition of the Maxwell multipoles} \label{Sec:MaxwellMultipolesDef}
\subsection{Definition through the Legendre polynomials}
There are several different ways to define the Maxwell multipoles. One particular definition starts from the observation that the two angles $\theta$ and $\varphi$ uniquely define a unit vector $ \mathbf{n}(\theta,\varphi)$, and vice versa. Therefore, we may choose to write the Maxwell multipoles in terms of $\mathbf{n}$, rather than $\theta$ and $\varphi$. We then define the Maxwell multipoles as those functions that only depend on $\mathbf{n}$, and are orthogonal under integration over the unit sphere.
This definition is very reminiscent of the Legendre polynomials $P_{\ell}(x)$, which are defined as those polynomials that only depend on a single variable $x$, and are orthogonal under integration over the interval $x \in [-1,1]$. To fix the prefactor, we choose the convention $P_{\ell}(1) = 1$.

To find explicit expressions for the Legendre polynomials, one can start from the following elemenary integral:
\begin{equation} \label{PolynomialIntegration}
\int_{-1}^1 x^\ell dx = \left\{ \begin{array}{ll}
\frac{2}{\ell + 1} & \text{with } \ell \text{ even} \\
0 & \text{with } \ell \text{ odd}
\end{array}
 \right.
\end{equation}
Then, a Gram-Schmidt orthogonalisation procedure gives the desired expressions. The first few are very well known:
\begin{align}
P_0(x) & = 1 \label{LP0} \\
P_1(x) & = x \label{LP1} \\
P_2(x) & = \frac{1}{2}(3x^2-1) \label{LP2} \\
P_3(x) & = \frac{1}{2}(5x^3-3x) \label{LP3}\\
P_4(x) & = \frac{1}{8} (35 x^4 - 30 x^2 + 3) \label{LP4} \\
& \vdots \nonumber
\end{align}
For example, it is now easy to verify that
\begin{equation}
    \int_{-1}^{1} P_1(x) P_3(x) dx = 0 ,
\end{equation}
by plugging in equations \eqref{LP1} and \eqref{LP3}, and evaluating the integrals using \eqref{PolynomialIntegration}.

The Maxwell multipoles can be constructed in exactly the same way. This is because there is a convenient formula for the angular integral over the product of $\ell$ unit vectors:
\begin{equation} \label{UnitVectorIntegration}
\int n_{i_1} n_{i_2} \ldots n_{i_\ell} d\Omega = \left\{ \begin{array}{ll}
\frac{4\pi}{\ell + 1} \delta_{(i_1 i_2} \delta_{i_3 i_4} \ldots \delta_{i_{\ell-1} i_{\ell})} & \text{with } \ell \text{ even} \\
0 & \text{with } \ell \text{ odd}
\end{array}
 \right.
\end{equation}
This equation can be found in the literature \cite{Thorne1980} but does not seem to be well-known. For completeness, we reproduce its derivation in appendix \ref{App:Int}. Equation \eqref{UnitVectorIntegration} is remarkably similar to \eqref{PolynomialIntegration}, with the only differences being the additional Kronecker deltas and a trivial factor $2\pi$. Since the integration rule \eqref{PolynomialIntegration} uniquely defines the Legendre polynomials, we might expect the following functions to be orthogonal under integration over the unit sphere:
\begin{align}
P^{(0)}(\mathbf{n}) & = 1 \label{MMP0} \\
P^{(1)}_i(\mathbf{n}) & = n_i \label{MMP1} \\
P^{(2)}_{ij}(\mathbf{n}) & = \frac{1}{2}(3n_i n_j-\delta_{ij}) \label{MMP2} \\
P^{(3)}_{ijk}(\mathbf{n}) & = \frac{1}{2}(5n_i n_j n_k-3n_{(i} \delta_{jk)}) \label{MMP3}\\
P^{(4)}_{ijkl}(\mathbf{n}) & = \frac{1}{8} (35 n_i n_j n_k n_l - 30 n_{(i} n_j \delta_{kl)} + 3 \delta_{(ij} \delta_{kl)}) \label{MMP4} \\
& \vdots \nonumber
\end{align}
We will take this as our definition of the Maxwell multipoles. We denote them as $P^{(\ell)}_{i_1 i_2 \ldots i_{\ell}}$ or $P_{(\ell)}^{i_1 i_2 \ldots i_{\ell}}$: the Maxwell multipole of order $\ell$ is a tensor of rank $\ell$. Notice the similarity between equations \eqref{MMP0}-\eqref{MMP4} and equations \eqref{LP0}-\eqref{LP4}: we obtain the Maxwell multipoles from the Legendre polynomials by replacing $x$ with a unit vector $\mathbf{n}$, adding Kronecker deltas until every term has $\ell$ indices, and then symmetrizing the resulting tensor.

It is insightful to explicitly write out the lowest order Maxwell multipoles in terms of $\theta$ and $\varphi$, by using equations \eqref{nxDef}-\eqref{nzDef}. They are:
\begin{equation}
P^{(0)}(\theta,\varphi) = 1
\end{equation}
\begin{equation}
P^{(1)}_i(\theta,\varphi) = 
\begin{pmatrix}
\sin\theta \cos\varphi \\
 \sin\theta \sin\varphi \\
\cos\theta
\end{pmatrix}
\end{equation}
\begin{equation}
P^{(2)}_{ij}(\theta,\varphi) = 
\begin{pmatrix}
\frac{1}{2} (3\sin^2\theta \cos^2\varphi - 1) & \frac{3}{2} \sin^2\theta \sin\varphi \cos\varphi & \frac{3}{2} \sin\theta \cos\theta \cos\varphi \\
\frac{3}{2} \sin^2\theta \sin\varphi \cos\varphi & \frac{1}{2} (3\sin^2\theta \sin^2\varphi - 1) & \frac{3}{2} \sin\theta \cos\theta \sin\varphi \\
\frac{3}{2} \sin\theta \cos\theta \cos\varphi & \frac{1}{2} \sin\theta \cos\theta \sin\varphi & \frac{3}{2}( 3\cos^2\theta - 1)
\end{pmatrix} 
\end{equation}
The orthogonality of the Maxwell multipoles would imply that all angular integrals of these components vanish if the components belong to different orders of $\ell$:
\begin{align}
\int P^{(\ell)}_{i_1 i_2 \ldots i_{\ell}}(\mathbf{n}) P^{(\ell')}_{j_1 j_2 \ldots j_{\ell'}}(\mathbf{n}) d\Omega & = 0 & \text{ if } \ell \neq \ell'
\end{align}
For example, taking the $x$-component from $\ell = 1$ and the $yy$-component from $\ell = 2$:
\begin{equation}
\int_{0}^{\pi} \int_0^{2\pi} \frac{1}{2} (3\sin^2\theta \sin^2\varphi - 1) \ \sin\theta \cos\varphi \ \sin\theta d\theta d\varphi = 0
\end{equation}
We have not yet proven that the Maxwell multipoles are orthogonal in general. In fact, they are, but we will postpone the proof of this statement to section \ref{Sec:Orthogonality}.

\subsection{Two other definitions}
The definition based on analogy with the Legendre polynomials is insightful but not very practical for proving theorems. In this section, we derive two other possible definitions of the Maxwell multipoles. The first is through a generating function:
\begin{equation} \label{DefGenerating}
\frac{1}{\sqrt{1-2\mathbf{n}\cdot\mathbf{q}+|\mathbf{q}|^2}} := \sum_{\ell = 0}^{+\infty} P^{(\ell)}_{i_1 i_2 \ldots i_{\ell}}(\mathbf{n}) q_{i_1} q_{i_2} \ldots q_{i_\ell}
\end{equation}
which is very similar to the generating function of the Legendre polynomials:
\begin{equation}
\frac{1}{\sqrt{1-2 x q+q^2}} := \sum_{\ell = 0}^{+\infty} P_{\ell}(x) q^\ell
\end{equation}
To see that \eqref{DefGenerating} is in fact the correct generating function, note that derivatives with respect to $q_i$ follow the same rules as derivatives with respect to a single variable $q$. The only difference is that $\frac{\partial q_i}{\partial q_j} = \delta_{ij}$, which gives rise to the Kronecker deltas in \eqref{MMP0}-\eqref{MMP4}. This generating function can also be found in the literature \cite{Efimov1979,Dennis2004} in other contexts. Notice that the generating function is the Coulomb potential of a charge placed on a position $\mathbf{n}$. It is therefore no surprise that the lowest order Maxwell multipoles are given by familiar formulas: \eqref{MMP1}-\eqref{MMP2} respectively represent the dipole moment vector and quadrupole moment matrix of a point charge in cartesian coordinates. 

A second alternative definition follows from the observation that the Maxwell multipoles are traceless symmetric tensors. Indeed, they are symmetric by construction, and it is easy to prove that they are traceless from the generating function:
\begin{align}
P^{(\ell)}_{j j i_1 i_2 \ldots i_{\ell-2}}(\mathbf{n}) & = \frac{1}{\ell !} \left. \frac{\partial^{\ell}}{\partial q_{i_1} \partial q_{i_2} \ldots \partial q_{i_{\ell-2}} \partial q_{j} q_{j}} \frac{1}{\sqrt{1-2 \mathbf{q} \cdot \mathbf{n} + q^2}} \right|_{\mathbf{q}=\mathbf{0}} \\
& = \frac{1}{\ell !} \left. \frac{\partial^{\ell-2}}{\partial q_{i_1} \partial q_{i_2} \ldots \partial q_{i_{\ell-2}}} \nabla^2 \frac{1}{|\mathbf{q}-\mathbf{n}|} \right|_{\mathbf{q}=\mathbf{0}} \\
& = 0
\end{align}
This means taking the traceless symmetric part of a Maxwell multipole leaves it unchanged:
\begin{equation}
P^{(\ell)}_{i_1 i_2 \ldots i_{\ell}} = P^{(\ell)}_{\{i_1 i_2 \ldots i_{\ell}\}}
\end{equation}
Comparing this with equations \eqref{MMP0}-\eqref{MMP4} and using \eqref{TracelessDelta}, we find another alternative definition of the Maxwell multipoles:
\begin{equation} \label{DefTraceless}
P^{(\ell)}_{i_1 i_2 \ldots i_{\ell}}(\mathbf{n}) =  \frac{1}{2^{\ell}} \binom{2\ell}{\ell} n_{\{ i_1} n_{i_2} \ldots n_{i_{\ell} \}}
\end{equation}
The prefactor $\frac{1}{2^{\ell}} \binom{2\ell}{\ell}$ is the leading coefficient of the Legendre polynomials. The definition of the Maxwell multipoles or similar functions as traceless symmetric tensors is a common one in the literature \cite{Thorne1980,Applequist1989,Zou2003,Copi2004}, although most authors define them without the prefactor. Several other authors \cite{Rubinstein2015,Ledesma2020} write the Maxwell multipoles as a product of the unit vectors and a position-independent traceless symmetric basis tensor $\mathcal{Y}_{i_1 i_2 \ldots i_{\ell}}=\mathcal{Y}_{\{i_1 i_2 \ldots i_{\ell}\}}$, which results in a scalar function:
\begin{equation} \label{DefScalar}
P(\mathbf{n}) = \mathcal{Y}_{i_1 i_2 \ldots i_{\ell}} n_{i_1} n_{i_2} \ldots n_{i_{\ell}}
\end{equation}
The tensor $\mathcal{Y}_{i_1 i_2 \ldots i_{\ell}}$ will be introduced in section \ref{Sec:SphericalHarmonics}. Here, we have shown that definitions \eqref{MMP0}-\eqref{MMP4}, \eqref{DefGenerating}, \eqref{DefTraceless}, \eqref{DefScalar} are all equivalent.

\subsection{Orthogonality of the Maxwell multipoles} \label{Sec:Orthogonality}
We now show that the Maxwell multipoles as defined by \eqref{MMP0}-\eqref{MMP4} are indeed orthogonal under integration over the unit sphere:
\begin{align}
\int P^{(\ell)}_{i_1 i_2 \ldots i_{\ell}}(\mathbf{n}) P^{(\ell')}_{j_1 j_2 \ldots j_{\ell'}}(\mathbf{n}) d\Omega = 0 & \hspace{10pt} \text{ if } \ell \neq \ell'
\end{align}
This statement is most easily proved by inserting the traceless symmetric form \eqref{DefTraceless} of the Maxwell multipoles. The resulting angular integral can be performed using equation \eqref{UnitVectorIntegration}:
\begin{small}
\begin{equation}
\int P^{(\ell)}_{i_1 i_2 \ldots i_{\ell}}(\mathbf{n}) P^{(\ell')}_{j_1 j_2 \ldots j_{\ell'}}(\mathbf{n}) d\Omega \nonumber = \frac{4 \pi}{\ell + \ell' + 1} \frac{1}{2^{\ell+\ell'}} \binom{2\ell}{\ell}\binom{2\ell'}{\ell'} \delta_{(\{ i_1 i_2} \ldots \delta_{i_{\ell-2}i_{\ell-1}} \delta_{i_\ell \} \{ j_1} \delta_{ j_2 j_3} \ldots \delta_{j_{\ell'-1} j_{\ell'}\} )}
\end{equation}
\end{small}
When performing the symmetrisation over the $\ell+\ell'$ indices in the last expression, all permutations that take two indices $i_1, i_2$ or $j_1, j_2$ under the same Kronecker delta will result in zero due to property \eqref{TracelessDelta}. Only one type of permutation remains, which is the one where every index $i$ is paired with an index $j$ and vice versa. This is only possible when $\ell = \ell'$, which completes the proof.

By counting the number of remaining permutations, the following explicit result can be obtained:
\begin{equation} \label{Orthogonality}
\int P^{(\ell)}_{i_1 i_2 \ldots i_{\ell}}(\mathbf{n}) P_{(\ell')}^{j_1 j_2 \ldots j_{\ell'}}(\mathbf{n}) d\Omega = \left\{ \begin{array}{lcl}
0 & \text{ if } & \ell \neq \ell' \\
\frac{4 \pi}{2\ell + 1} \frac{1}{2^{\ell}} \binom{2\ell}{\ell} \delta_{i_1}^{\{ j_1} \delta_{i_2}^{j_2} \ldots \delta_{i_\ell}^{j_\ell \}} & \text{ if } & \ell = \ell'
\end{array} \right.
\end{equation}
This result is similar to the orthogonality theorem for the Legendre polynomials, with the main difference being the additional Kronecker deltas. These ensure that the independent traceless components of the Maxwell multipoles are orthogonal as well. As we shall see in section \ref{Sec:SphericalHarmonics}, the orthogonality theorem for the spherical harmonics can be derived from \eqref{Orthogonality}. We stress that $\delta_{i}^{j} = \delta_{ij}$ since we make no distinction between upper and lower indices: we only use lower and upper indices to group the different indices together.

In additional to being orthogonal, one can also prove that the Maxwell multipoles form a complete basis for functions over the unit sphere \cite{Sylvester1876}:
\begin{equation} \label{Completeness}
\sum_{\ell = 0}^{+\infty} \frac{2 \ell + 1}{4\pi} \frac{2^\ell}{\binom{2\ell}{\ell}} P^{(\ell)}_{i_1 i_2 \ldots i_{\ell}}(\mathbf{n}) P^{(\ell)}_{i_1 i_2 \ldots i_{\ell}}(\mathbf{s}) = \delta(\mathbf{n}-\mathbf{s}),
\end{equation}
which is rather elementary to prove using the properties of the next section. This property guarantees that we can make a multipole expansion of an arbitrary function in terms of the functions $P^{(\ell)}_{i_1 i_2 \ldots i_{\ell}}(\mathbf{n})$.

\section{Properties} \label{Sec:Properties}
\begin{table}
\begin{tabular}{|c|c|}
\hline
\multicolumn{2}{|c|}{\textbf{Properties of the Maxwell multipoles $P^{(\ell)}_{i_1 i_2 \ldots i_{\ell}}(\mathbf{n})$}} \\ \hline
Multipole expansion & $ \begin{array}{rl}
\displaystyle f(\mathbf{n}) & := \displaystyle \sum_{\ell = 0}^{+\infty} f^{(\ell)}_{i_1 i_2 \ldots i_{\ell}} P^{(\ell)}_{i_1 i_2 \ldots i_{\ell}}(\mathbf{n}) \\
\displaystyle f^{(\ell)}_{i_1 i_2 \ldots i_{\ell}} & =\displaystyle \frac{2 \ell + 1}{4\pi} \frac{2^\ell}{\binom{2\ell}{\ell}}  \int P^{(\ell)}_{i_1 i_2 \ldots i_{\ell}}(\mathbf{n}) f(\mathbf{n}) d\Omega \\
\displaystyle \int f(\mathbf{n}) g(\mathbf{n}) d\Omega & = \displaystyle \sum_{\ell = 0}^{+\infty} \frac{4 \pi}{2 \ell + 1} \frac{1}{2^{\ell}} \binom{2\ell}{\ell} f^{(\ell)}_{i_1 i_2 \ldots i_{\ell}} g^{(\ell)}_{i_1 i_2 \ldots i_{\ell}}
\end{array}$ \\ \hline
Orthogonality & $\begin{array}{rl}
\displaystyle \int P^{(\ell)}_{i_1 i_2 \ldots i_{\ell}}(\mathbf{n}) P_{(\ell')}^{j_1 j_2 \ldots j_{\ell'}}(\mathbf{n}) d\Omega & = \displaystyle \delta_{\ell,\ell'} \frac{4 \pi}{2\ell + 1} \frac{1}{2^{\ell}} \binom{2\ell}{\ell} \delta_{i_1}^{\{ j_1} \delta_{i_2}^{j_2} \ldots \delta_{i_\ell}^{j_\ell \}} \\
\displaystyle  \int P^{(\ell)}_{i_1 i_2 \ldots i_{\ell}}(\mathbf{n}) n_{j_1} n_{j_2} \ldots n_{j_{\ell'}}  d\Omega & = 0 \hspace{10pt} \text{ if } \ell' < \ell
\end{array}$ \\ \hline
Recurrence relations & $\begin{array}{rl}
\displaystyle (2\ell + 1) n^{\phantom{(}}_{j} P^{(\ell)}_{i_1 i_2 \ldots i_{\ell}}(\mathbf{n}) & = \displaystyle (\ell + 1) P^{{(\ell+1)}}_{j i_1 i_2 \ldots i_{\ell}}(\mathbf{n}) + (2\ell-1) \delta^{\phantom{(}}_{j \{i_1} P^{{(\ell-1)}}_{i_2 i_3 \ldots i_{\ell} \}}(\mathbf{n}) \\
\displaystyle n^{\phantom{(}}_{j} P^{(\ell)}_{j i_1 i_2 \ldots i_{\ell-1}}(\mathbf{n}) & =\displaystyle  P^{{(\ell-1)}}_{i_1 i_2 \ldots i_{\ell-1}}(\mathbf{n}) \end{array}$ \\ \hline
Link to $P_{\ell}(x)$ & $ \begin{array}{rl}
\displaystyle s_{i_1} s_{i_2} \ldots s_{i_\ell} P^{(\ell)}_{i_1 i_2 \ldots i_{\ell}}(\mathbf{n}) & = \displaystyle P_{\ell}(\mathbf{n} \cdot \mathbf{s}) \\
\displaystyle P^{(\ell)}_{i_1 i_2 \ldots i_{\ell}}(\mathbf{s}) P^{(\ell)}_{i_1 i_2 \ldots i_{\ell}}(\mathbf{n}) & = \displaystyle \frac{1}{2^{\ell}} \binom{2l}{l} P_{\ell}(\mathbf{n} \cdot \mathbf{s}) \end{array}$ \\ \hline
Angular integrals & $\begin{array}{rl}
\displaystyle \int P^{(\ell)}_{i_1 i_2 \ldots i_{\ell}}(\mathbf{n}) f(\mathbf{n}\cdot\mathbf{s}) d\Omega & = 2\pi \displaystyle P^{(\ell)}_{i_1 i_2 \ldots i_{\ell}}(\mathbf{s}) \int_{-1}^1 P_{\ell}(x) f(x) dx
\end{array} $ \\ \hline
Rotation transformations & $\begin{array}{rl}
\displaystyle P^{(\ell)}_{i_1 i_2 \ldots i_\ell}(\mathcal{S} \cdot \mathbf{n}) & = \displaystyle  S^{\phantom{(}}_{i_1 j_1} S^{\phantom{(}}_{i_2 j_2} \ldots S^{\phantom{(}}_{i_{\ell} j_{\ell}} P^{(\ell)}_{j_1 j_2 \ldots j_\ell}(\mathbf{n})
\end{array}$ \\ \hline
Laplacian eigenfunctions & $\begin{array}{rl}
\displaystyle r^2 \nabla^2 P^{(\ell)}_{i_1 i_2 \ldots i_{\ell}}(\mathbf{n}) & = \displaystyle -\ell(\ell+1) P^{(\ell)}_{i_1 i_2 \ldots i_{\ell}}(\mathbf{n}) \end{array} $ \\ \hline
Traceless & $\begin{array}{rl}
\displaystyle P^{(\ell)}_{jj i_1 i_2 \ldots i_{\ell-2}}(\mathbf{n}) & = 0 \end{array} $ \\ \hline
\end{tabular}
\caption{\label{Tab:Properties} A summary of the most useful properties of the Maxwell multipoles. In these expressions, $\mathbf{n} = \mathbf{n}(\theta,\varphi)$ as defined by \eqref{nxDef}-\eqref{nzDef}, $\mathbf{s}$ is a fixed unit vector, $d\Omega := \sin \theta d\theta d\varphi$, $f(x)$ is an arbitrary continuous function, and $\mathcal{S}$ is an orthogonal rotation matrix.}
\end{table}

From the definitions \eqref{DefGenerating} and \eqref{DefTraceless}, many useful properties of the Maxwell multipoles may be derived. These properties are listed in table \ref{Tab:Properties}. Many of these properties are similar to those of the Legendre polynomials, and can be proven from the generating function \eqref{DefGenerating} or the traceless symmetric definition \eqref{DefTraceless} in a very similar and straightforward way.

Rather than presenting proofs for the theorems in table \ref{Tab:Properties}, which is a good excercise to get acquainted with the Maxwell multipoles and their notation, we present an example of how these theorems are used in practice to calculate integrals. Consider the potential generated by a quadrupole (in Gaussian units):
\begin{equation}
V(\mathbf{r}) = \frac{1}{r^3} \mathbf{n}\cdot\mathcal{Q}\cdot\mathbf{n}
\end{equation}
where the quadrupole moment $\mathcal{Q}$ is a traceless $3 \times 3$ matrix. Here, we calculate the Fourier transform of this potential. From \eqref{MMP0} and \eqref{MMP2}, and using that $\mathcal{Q}$ is traceless, we can immediately see that:
\begin{equation}
V(\mathbf{r}) =\frac{2}{3r^3} Q_{ij} P^{(2)}_{ij}(\mathbf{n})
\end{equation}
The Fourier transform can then be calculated as follows, where $\mathbf{s} :=  \mathbf{k}/k$:
\begin{align}
\tilde{V}(\mathbf{k}) & = \int f(\mathbf{r}) e^{i \mathbf{k}\cdot \mathbf{r}} d^3\mathbf{r} \\
& = \frac{2}{3} A_{ij} \int_0^{+\infty} \frac{1}{r} \left( \int P^{(2)}_{ij}(\mathbf{n})e^{i kr \mathbf{n}\cdot\mathbf{s}} d\Omega \right) dr
\end{align}
The angular integral is of the form presented in table \ref{Tab:Properties}:
\begin{align}
\tilde{V}(\mathbf{k}) & = \frac{4\pi}{3} Q_{ij} P^{(2)}_{ij}(\mathbf{s})\int_0^{+\infty} \int_{-1}^{1} \frac{e^{i kr x}}{r} P_{2}(x) dr dx \\
& = \pi \mathbf{s}\cdot \mathcal{Q} \cdot \mathbf{s} \int_0^{+\infty} \int_{-1}^{1} \frac{e^{i kr x}}{r} (3x^2-1) dr dx
\end{align}
The remaining integrals are elementary, and we obtain:
\begin{equation}
\tilde{V}(\mathbf{k}) = -\frac{4\pi}{3} \frac{\mathbf{k}\cdot \mathcal{Q} \cdot \mathbf{k}}{k^2}
\end{equation}
The beauty of this calculation is that it is done entirely in vectorized form. If we were to make a multipole expansion in terms of spherical harmonics, we would have to write out the scalar products explicitly:
\begin{equation}
V(\mathbf{r}) = \frac{1}{r^3} \left(Q_{xx} \sin^2\theta \cos^2\varphi + (Q_{xy}+Q_{yx}) \sin^2\theta \sin\varphi \cos\varphi + Q_{yy} \sin^2\theta \cos^2\varphi + \ldots \right)
\end{equation}
since the quadrupole tensor is in its cartesian form. The results would be the same in the end, but it is clear that the calculation using Maxwell multipoles is a lot less tedious. The function $V(\mathbf{r})$ in this example was quite simple. In other physical problems, one usually encounters functions that are much more complicated \cite{Houtput2021}. Still, it is usually possible to calculate the multipole expansion coefficients $f^{(\ell)}_{i_1 i_2 \ldots i_{\ell}}$ by combining the different properties in \ref{Tab:Properties}, without directly calculating an angular integral.

\section{Connection to the spherical harmonics} \label{Sec:SphericalHarmonics}
From equation \eqref{Completeness}, it is clear that the Maxwell multipoles form a complete basis for functions on the unit sphere. In addition, it can be shown (see table \ref{Tab:Properties}) that they are eigenfunctions of $r^2 \nabla^2$, the angular part of the Laplacian in spherical coordinates, with eigenvalues $-\ell(\ell+1)$. This means there must be a 1-to-1 correspondence between the Maxwell multipoles and the spherical harmonics $Y_{\ell,m}(\theta,\varphi)$, since they also satisfy these properties. In this section, we investigate this correspondence and give explicit formulas to convert from Maxwell multipoles to spherical harmonics, and vice versa. 

First, note that a traceless symmetric tensor with $\ell$ indices has $2\ell+1$ independent components. This is precisely the amount of spherical harmonics belonging to the eigenvalue $\ell$: they are indexed with the quantum number $m$. Thus, for fixed $\ell$, we can associate the spherical harmonics $Y_{\ell,m}(\theta,\varphi)$ with the different independent components of the Maxwell multipole $P^{(\ell)}_{i_1 i_2 \ldots i_{\ell}}(\mathbf{n})$. All that remains is to find $(2\ell+1)$ basis tensors for traceless symmetric tensors of rank $\ell$. To find these tensors, we use the following property. If we define the vectors $\mathbf{u}^{(-1)}$, $\mathbf{u}^{(0)}$, and $\mathbf{u}^{(1)}$ as follows:
\begin{align}
\mathbf{u}^{(-1)} & := \begin{pmatrix}
1 \\
i \\
0
\end{pmatrix},&
\mathbf{u}^{(0)} & := \begin{pmatrix}
0 \\
0 \\
1
\end{pmatrix},&
\mathbf{u}^{(1)} & := \begin{pmatrix}
-1 \\
i \\
0
\end{pmatrix}
\label{uDef}
\end{align}
then it is straightforward to show from equations \eqref{MMP0}-\eqref{MMP4} that \cite{Thorne1980,Ledesma2020}:
\begin{equation} \label{MaxwellSphericalLink}
P^{(\ell)}_{i_1 i_2 \ldots i_{\ell}}(\mathbf{n}) u^{(m_1)}_{i_1} u^{(m_2)}_{i_2} \ldots u^{(m_\ell)}_{i_\ell} = (-1)^m \sqrt{\frac{4\pi}{2\ell+1}} \frac{\sqrt{(\ell-m)!(\ell+m)!}}{l!} Y_{\ell,-m}(\theta,\varphi)
\end{equation}
where on the left hand side $m_j \in \{-1,0,1\}$ and on the right hand side $\displaystyle m = \sum_{j=1}^{\ell} m_j$. The result is quite intuitive: $\mathbf{u}^{(-1)}\cdot\mathbf{n} \sim e^{i\varphi}$, so this adds angular momentum to the function. Meanwhile $\mathbf{u}^{(1)}\cdot\mathbf{n} \sim e^{-i\varphi}$ so it subtracts angular momentum. Therefore, we can define the following basis tensors, labelled by an additional index $m$:
\begin{equation} \label{BasisDef}
\mathcal{Y}^{(\ell,m)}_{i_1 i_2 \ldots i_{\ell}} := \sqrt{\frac{1}{2^{\ell}} \binom{2\ell}{\ell-m}} \times \left\{ \begin{array}{ll}
u^{(1)}_{\{i_1} u^{(1)}_{i_2} \ldots u^{(1)}_{i_m} u^{(0)}_{i_{m+1}} u^{(0)}_{i_{m+2}} \ldots u^{(0)}_{i_\ell\}} & \text{ if } m \geq 0 \\
u^{(-1)}_{\{i_1} u^{(-1)}_{i_2} \ldots u^{(-1)}_{i_{|m|}} u^{(0)}_{i_{|m|+1}} u^{(0)}_{i_{|m|+2}} \ldots u^{(0)}_{i_\ell\}} & \text{ if } m \leq 0
\end{array}
\right.
\end{equation}
These basis tensors are orthonormal and complete:
\begin{align}
\mathcal{Y}^{(\ell,m)*}_{i_1 i_2 \ldots i_{\ell}} \mathcal{Y}^{(\ell,m')}_{i_1 i_2 \ldots i_{\ell}} & = \delta_{m,m'} \label{BasisOrthonormal} \\
\sum_{m=-\ell}^{\ell} \mathcal{Y}^{(\ell,m)*}_{i_1 i_2 \ldots i_{\ell}} \mathcal{Y}^{(\ell,m)}_{j_1 j_2 \ldots j_{\ell}} & = \delta_{i_1}^{\{ j_1} \delta_{i_2}^{j_2} \ldots \delta_{i_\ell}^{j_\ell \}} \label{BasisComplete}
\end{align}
and it holds that $\mathcal{Y}^{(\ell,m)*} = (-1)^m \mathcal{Y}^{(\ell,-m)}$. The first few of these basis tensors are:
\begin{align}
&& \mathcal{Y}^{(0,0)} & = 1 && \\
\mathcal{Y}^{(1,-1)}_i & := \begin{pmatrix}
\frac{1}{\sqrt{2}} \\
\frac{i}{\sqrt{2}}  \\
0
\end{pmatrix},&
\mathcal{Y}^{(1,0)}_i & := \begin{pmatrix}
0 \\
0 \\
1
\end{pmatrix},&
\mathcal{Y}^{(1,1)}_i & := \begin{pmatrix}
\frac{-1}{\sqrt{2}}  \\
\frac{i}{\sqrt{2}}  \\
0
\end{pmatrix} \\
\mathcal{Y}^{(2,-1)}_{ij} & := \begin{pmatrix}
0 & 0 & \frac{1}{2} \\
0 & 0 & \frac{i}{2} \\
\frac{1}{2} & \frac{i}{2} & 0
\end{pmatrix},&
\mathcal{Y}^{(2,0)}_{ij} & := \begin{pmatrix}
-\frac{1}{\sqrt{6}} & 0 & 0 \\
0 & -\frac{1}{\sqrt{6}} & 0 \\
0 & 0 & \frac{2}{\sqrt{6}}
\end{pmatrix},&
\mathcal{Y}^{(2,1)}_{ij} & := \begin{pmatrix}
0 & 0 & -\frac{1}{2} \\
0 & 0 & \frac{i}{2} \\
-\frac{1}{2} & \frac{i}{2} & 0
\end{pmatrix},& \\
\mathcal{Y}^{(2,-2)}_{ij} & := \begin{pmatrix}
\frac{1}{2} & \frac{i}{2} & 0 \\
\frac{i}{2} & -\frac{1}{2} & 0 \\
0 & 0 & 0
\end{pmatrix},& &&
\mathcal{Y}^{(2,2)}_{ij} & := \begin{pmatrix}
\frac{1}{2} & -\frac{i}{2} & 0 \\
-\frac{i}{2} & -\frac{1}{2} & 0 \\
0 & 0 & 0
\end{pmatrix},&
\end{align}
Since this basis is complete \eqref{BasisComplete}, any traceless symmetric tensor can be decomposed in this basis. From equations \eqref{MaxwellSphericalLink}-\eqref{BasisDef}, we see that if we decompose the Maxwell mulipoles in this basis, the independent components will be the spherical harmonics:
\begin{align}
P^{(\ell)}_{i_1 i_2 \ldots i_{\ell}}(\mathbf{n}) & =  \sqrt{\frac{4\pi}{2\ell+1}\frac{1}{2^{\ell}}\binom{2\ell}{\ell}} \sum_{m=-\ell}^{\ell} Y_{\ell,m}(\theta,\varphi) \mathcal{Y}^{(\ell,m)}_{i_1 i_2 \ldots i_{\ell}} \label{YlmMaxwellLink1} \\
Y_{\ell,m}(\theta,\varphi) & = \sqrt{\frac{2\ell+1}{4\pi}\frac{2^{\ell}}{\binom{2\ell}{\ell}}} P^{(\ell)}_{i_1 i_2 \ldots i_{\ell}}(\mathbf{n}) \mathcal{Y}^{(\ell,m)*}_{i_1 i_2 \ldots i_{\ell}}
 \label{YlmMaxwellLink2}
\end{align}

\begin{table}
\begin{tabular}{|c|c|}
\hline
\multicolumn{2}{|c|}{\textbf{Properties of the general spherical harmonics $\tilde{Y}_{\ell,m}(\theta,\varphi)$}} \\ \hline
Multipole expansion & $ \begin{array}{rl}
\displaystyle f(\theta,\varphi) & := \displaystyle \sum_{\ell,m} f_{\ell,m} \tilde{Y}_{\ell,m}(\theta,\varphi) \\
\displaystyle f_{\ell,m} & =\displaystyle \int \tilde{Y}^*_{\ell,m}(\theta,\varphi) f(\theta,\varphi) d\Omega \\
&\displaystyle = \sqrt{\frac{4\pi}{2\ell+1} \frac{1}{2^{\ell}}\binom{2\ell}{\ell}} \mathcal{\tilde{Y}}^{(\ell,m)}_{i_1 i_2 \ldots i_{\ell}} f^{(\ell)}_{i_1 i_2 \ldots i_{\ell}} \\
\displaystyle \int f(\theta,\varphi) g(\theta,\varphi) d\Omega & = \displaystyle \sum_{\ell,m} f^*_{\ell,m} g^{\phantom{*}}_{\ell,m}
\end{array}$ \\ \hline
Orthogonality & $\begin{array}{rl}
\displaystyle \int \tilde{Y}^*_{\ell,m}(\theta,\varphi)\tilde{Y}_{\ell',m'}(\theta,\varphi) d\Omega & = \displaystyle \delta_{\ell,\ell'} \delta_{m,m'} \\
\displaystyle  \int \tilde{Y}_{\ell,m}(\theta,\varphi) n_{j_1} n_{j_2} \ldots n_{j_{\ell'}}  d\Omega & = 0 \hspace{10pt} \text{ if } \ell' < \ell
\end{array}$ \\ \hline
Link to $P_{\ell}(x)$ & $ \displaystyle \frac{4\pi}{2\ell+1} \sum_{m=-\ell}^{\ell} \tilde{Y}_{\ell,m}^*(\mathbf{n}) \tilde{Y}_{\ell,m}(\mathbf{s}) = P_{\ell}(\mathbf{n}\cdot\mathbf{s})$ \\ \hline
Angular integrals & $ \displaystyle \int \tilde{Y}_{\ell,m}(\mathbf{n}) f(\mathbf{n}\cdot\mathbf{s}) d\Omega = 2\pi \tilde{Y}_{\ell,m}(\mathbf{s}) \int_{-1}^{1} P_{\ell}(x) f(x) dx $ \\ \hline
\end{tabular}
\caption{\label{Tab:YlmProperties} A summary of the most useful properties of the general spherical harmonics, derived from the properties of table \ref{Tab:Properties}. In these formulas, $\tilde{Y}_{\ell,m}(\theta,\varphi)$ represent any set of independent components of the Maxwell mulipoles, and $\tilde{Y}_{\ell,m}(\mathbf{n})=\tilde{Y}_{\ell,m}(\mathbf{n}(\theta,\varphi))$.}
\end{table}

From equation \eqref{YlmMaxwellLink1} and the properties of table \ref{Tab:Properties}, it is possible to derive the familiar properties of the spherical harmonics. For example, from the orthogonality theorem \eqref{Orthogonality}, it can be easily seen that the independent components $Y_{\ell,m}(\theta,\varphi)$ must also be orthogonal:
\begin{equation} \label{YlmOrthogonality}
\int Y^*_{\ell,m}(\theta,\varphi) Y_{\ell',m'}(\theta,\varphi) d\Omega = \delta_{\ell,\ell'} \delta_{m,m'}
\end{equation}
which is the well-known orthogonality theorem of the spherical harmonics. However, we note that the basis tensors \eqref{BasisDef} are not unique: any orthonormal basis instead of \eqref{BasisDef} will lead to a different set of spherical harmonics $\tilde{Y}_{\ell,m}(\theta,\varphi)$. that satisfy the orthogonality relation \eqref{YlmOrthogonality}. For example, we may take the real and imaginary parts of \eqref{BasisDef} to define a different orthonormal basis, which will lead to the real spherical harmonics (e.g. \cite{Blanco1997}). 
The orthogonality of the real spherical harmonics is also guaranteed by \eqref{Orthogonality}.

In general, any set of orthonormal basis tensors $\tilde{\mathcal{Y}}^{(\ell,m)}_{i_1 i_2 \ldots i_{\ell}}$ will lead to a set of orthogonal spherical harmonics $\tilde{Y}_{\ell,m}(\theta,\varphi)$. The properties of these new functions follow directly from the properties of the Maxwell multipoles in table \ref{Tab:Properties}. We present these properties in table \ref{Tab:YlmProperties}. For the standard spherical harmonics $Y_{\ell,m}(\theta,\varphi)$, all these properties are well known, but it is not immediately obvious that they should hold for any set of independent components of the Maxwell multipoles.

We note that the decomposition presented in this section is also useful from a numerical viewpoint. Maxwell multipoles are typically a tool for analytic calculations, since storing an $\ell$-dimensional tensor quickly becomes impractical for large $\ell$. Therefore, it is important that at any point during the calculation, the Maxwell multipoles can be converted to the more standard spherical harmonics, whose multipole expansion coefficients are stored with ease.

\section{Summary} \label{Sec:Conclusions}

In this article, we introduce the Maxwell multipoles in an intuitive way. They are quite similar to the spherical harmonics and are an excellent tool for making a multipole expansion or calculating an angular integral. Usually, one can avoid doing an explicit angular integral and simply use the properties from section \ref{Sec:Properties}. However, it is important to use the correct functions for the job. A good rule of thumb is that the spherical harmonics are good for problems with scalar expressions, while the Maxwell multipoles are useful for any expressions in vectorized or tensorial form. Consider, for example, the following two integrals:
\begin{align}
 & \int \frac{\sin^2\theta \cos\theta \cos^2\varphi}{|\mathbf{r}-\mathbf{a}|} d\Omega \\
 & \int \frac{A_{ijk} n_i n_j n_k}{|\mathbf{r}-\mathbf{a}|} d\Omega
\end{align}
Both can be solved using a multipole expansion, as shown in section \ref{Sec:Properties}. For the first integral, we can simply expand in spherical harmonics: tensor notation is unnecessarily complex in this context. However, for the second integral, we do not want to consider the 10 separate components of the tensor $A_{ijk}$, and an expansion in Maxwell multipoles is preferred.
At any time, we can convert from the Maxwell multipoles to the spherical harmonics and vice versa by using the basis tensors defined in section \ref{Sec:SphericalHarmonics}.

The theory and properties surrounding Maxwell multipoles is not presented in a unified way in the current literature, mostly due to the different conventions and names that different authors use. We circumvented this problem by introducing the Maxwell multipoles from scratch, based on the conceptually simpler Legendre polynomials. We have shown that this definition is equivalent to the various other definitions found in the literature \cite{Bruno2018, Thorne1980, Zou2003, Dennis2004, Efimov1979, Applequist1989, Copi2004, Rubinstein2015}. We have presented the reader with a practical reference for the use of Maxwell multipoles, and listed their most important properties in one location. The method presented in this article greatly simplifies calculations when a multipole expansion or angular integral of a function in vectorized form must be calculated.

\acknowledgments

This research was funded by the University Research Fund (BOF) of the University of Antwerp.

\appendix

\section{Symmetric detracing operator} \label{App:detracing}
In this appendix, we present the general formula for the symmetric detracing of arbitrary tensors. Suppose we want to calculate $T_{\{i_1 i_2 \ldots i_{\ell}\}}$, where each of the indices can take on $n$ values. We start by constructing the symmetric part of this tensor, which we will denote with $S$:
\begin{equation}
S_{i_1 i_2 \ldots i_{\ell}}:=T_{(i_1 i_2 \ldots i_{\ell})}
\end{equation}
Then, the traceless symmetric part can be defined as:
\begin{align}
& T_{\{ i_1 i_2 \ldots i_{\ell} \}} = \\
& \sum_{k=0}^{\floor{\frac{\ell}{2}}} \frac{(-1)^k}{2^{2k}} \frac{\Gamma(\ell+1) \Gamma\left(\frac{n}{2}+\ell-k-1\right)}{\Gamma(\ell-2k+1) \Gamma(k+1) \Gamma\left(\frac{n}{2}+\ell-1\right)} S_{j_1 j_1 j_2 j_2 \ldots j_k j_k (i_{1} i_{2} \ldots i_{\ell-2k}} \delta_{i_{\ell-2k+1} i_{\ell-2k+2}} \ldots \delta_{i_{\ell-1} i_{\ell})} 
\end{align}
For the scope of this article, we only require the specific case $n=3$. In that case, the formula reduces to:
\begin{equation}
T_{\{ i_1 i_2 \ldots i_{\ell} \}} = \frac{1}{\binom{2\ell}{\ell}} \sum_{k=0}^{\floor{\frac{\ell}{2}}} (-1)^k \binom{\ell}{k} \binom{2\ell-2k}{\ell} S_{j_1 j_1 j_2 j_2 \ldots j_k j_k (i_{1} i_{2} \ldots i_{\ell-2k}} \delta_{i_{\ell-2k+1} i_{\ell-2k+2}} \ldots \delta_{i_{\ell-1} i_{\ell})} 
\end{equation}
It is quite remarkable that these coefficients are the same as those in the definition of the Legendre polynomials, up to a constant prefactor $\frac{1}{2^{\ell}} \binom{2\ell}{\ell}$. This is why the Maxwell multipoles can be defined in two different ways: the standard definition as traceless symmetric tensors, and definition \eqref{MMP0}-\eqref{MMP4} by direct analogy with the Legendre polynomials.

\section{Calculation of the basic angular integral} \label{App:Int}
The entire article is based on formula \eqref{UnitVectorIntegration}, for the angular integral over $\ell$ identical unit vectors \cite{Thorne1980}:
\begin{equation} \label{AppFormula}
\mathcal{I} := \int n_{i_1} n_{i_2} \ldots n_{i_\ell} d\Omega = \left\{ \begin{array}{ll}
\frac{4\pi}{\ell + 1} \delta_{(i_1 i_2} \delta_{i_3 i_4} \ldots \delta_{i_{\ell-1} i_{\ell})} & \text{with } \ell \text{ even} \\
0 & \text{with } \ell \text{ odd}
\end{array}
 \right.
\end{equation}
In reference \cite{Thorne1980}, there is no proof for this formula, and since we did not find any proof elsewhere, we produce one here.

We may use the definition \eqref{nxDef}-\eqref{nzDef} of the unit vector and $d\Omega := \sin\theta d\theta d\varphi$ to calculate the integral.
Since \eqref{AppFormula} is symmetric in the indices $i_1 i_2 \ldots i_\ell$, we can also write it as follows:
\begin{equation}
\mathcal{I} := \int n_{i_1} n_{i_2} \ldots n_{i_\ell} d\Omega = \delta_{(i_1}^{j_1} \delta_{i_2}^{j_2} \ldots \delta_{i_\ell)}^{j_\ell} \int n_{j_1} n_{j_2} \ldots n_{j_{\ell}} d\Omega
\end{equation}
where we again emphasize that we make no distinction between lower and upper indices.

By symmetry, the integral will be $0$ unless each of the indices $x$, $y$ and $z$ appear an even number of times. This already proves that the integral is $0$ if $\ell$ is odd. For the remainder of the proof, we can assume that $\ell := 2M$ with $M \in \mathbb{N}$.

The proof proceeds by explicitly performing the summation over the indices $j_1 j_2 \ldots j_\ell$. Every term in the summation only depends on the amount of times the indices $j_1 j_2 \ldots j_{\ell}$ take on the values $x$, $y$ and $z$. Call these amounts $2m$, $2(n-m)$ and $2M-2n$ respectively. Then, the integral can be written as: 
\begin{align}
\mathcal{I} = & \sum_{n=0}^{M} \sum_{m=0}^{n} \binom{2M}{2m,2n-2m,2M-2n} \delta_{(i_1}^{x} \ldots \delta_{i_{2m}}^{x} \delta_{i_{2m+1}}^{y} \ldots \delta_{i_{2n}}^{y} \delta_{i_{2n+1}}^{z} \ldots \delta_{i_{2M})}^{z} \times \nonumber \\
 & \ \times  \int_0^{\pi} \sin^{2n+1}(\theta)\cos^{2M-2n}(\theta) d\theta \int_0^{2\pi} \cos^{2m}(\varphi) \sin^{2(n-m)}(\varphi) d\varphi \label{AppA_int1}
\end{align}
where the multinomial coefficient $\binom{2M}{2m,2n,2M-2n}$ indicates the amount of possible combinations that leads to the desired term. The integrals over $\varphi$ and $\theta$ are standard integrals that can be written in terms of the Gamma function:
\begin{align}
\int_0^{2\pi} \cos^{2m}(\varphi) \sin^{2n-2m}(\varphi) d\varphi & = \frac{4 \pi}{2^{2n-1}} \frac{\Gamma(2n-2m) \Gamma(2m)}{\Gamma(n-m) \Gamma(m) \Gamma(n+1)} \\
\int_0^{\pi} \sin^{2n+1}(\theta)\cos^{2M-2n}(\theta) d\theta & = 2^{2n+2} \frac{\Gamma(n+1) \Gamma(M+1) \Gamma(2M-2n)}{\Gamma(2M+1) \Gamma(M-n)}
\end{align}
Using this result in (\ref{AppA_int1}), a straightforward calculation then leads to:
\begin{equation}
\mathcal{I} =  \frac{4\pi}{2M+1} \sum_{m+n+k=M} \binom{M}{m,n,k} \delta_{(i_1}^{x} \ldots \delta_{i_{2m}}^{x} \delta_{i_{2m+1}}^{y} \ldots \delta_{i_{2m+2n}}^{y} \delta_{i_{2m+2n+1}}^{z} \ldots \delta_{i_{2M})}^{z}
\end{equation}
This is a multinomial expansion, which is equivalent to:
\begin{equation}
\mathcal{I} = \frac{4\pi}{2M+1} \left(\delta_{(i_1}^x \delta_{i_2}^x + \delta_{i_1}^y \delta_{i_2}^y + \delta_{i_1}^z \delta_{i_2}^z \right) \ldots \left(\delta_{i_{2M-1}}^x \delta_{i_{2M}}^x + \delta_{i_{2M-1}}^y \delta_{i_{2M}}^y + \delta_{i_{2M-1}}^z \delta_{i_{2M})}^z \right)
\end{equation}
Finally, by noting that
\begin{equation}
\delta_{i}^x \delta_{j}^x + \delta_{i}^y \delta_{j}^y + \delta_{i}^z \delta_{j}^z = \delta_{i j}
\end{equation}
and going back to $2M = \ell$, we can write this result as:
\begin{equation} \label{AppUnitIntRes}
\int n_{i_1} n_{i_2} \ldots n_{i_\ell} d\Omega = \left\{ \begin{array}{ll}
\frac{4\pi}{\ell + 1} \delta_{(i_1 i_2} \delta_{i_3 i_4} \ldots \delta_{i_{\ell-1} i_{\ell})} & \text{with } \ell \text{ even} \\
0 & \text{with } \ell \text{ odd}
\end{array}
 \right.
\end{equation}
This is exactly expression \eqref{AppA_int1}, which completes the proof.

\bibliography{References}

\end{document}